\documentclass{ws-procs9x6}
\usepackage{rotating_pr}

\def\babar{\mbox{\sl B\hspace{-0.35em} {\footnotesize\sl A}\hspace{-0.35em}
B\hspace{-0.35em} {\footnotesize\sl A\hspace{-0.1em}R}}}
\def\ra{\rightarrow}

\def\lf    {\mathrm{LF}}
\def\ifb   {\mathrm{fb^{-1}}}
\def\tt    {\mathrm{\tau^{+}\tau^{-}}}
\def\mm    {\mathrm{\mu^{+}\mu^{-}}}
\def\qq    {\mathrm{q\bar{q}}}

\def\epem  {\mathrm{e^+e^-}}

\def\GeV{\ifmmode{\mathrm{Ge\kern -0.1em V}}\else
                  \textrm{Ge\kern -0.1em V}\fi}
\def\MeV{\ifmmode{\mathrm{Me\kern -0.1em V}}\else
                  \textrm{Me\kern -0.1em V}\fi}

\newcommand{\eee}     {\ensuremath{e^-\!e^+\!e^-}}
\newcommand{\eemw}    {\ensuremath{\mu^+\!e^-\!e^-}}
\newcommand{\eemr}    {\ensuremath{\mu^-\!e^+\!e^-}}
\newcommand{\emmw}    {\ensuremath{e^+\!\mu^-\!\mu^-}}
\newcommand{\emmr}    {\ensuremath{e^-\!\mu^+\!\mu^-}}
\newcommand{\mmm}     {\ensuremath{\mu^-\!\mu^+\!\mu^-}}

\newcommand{\Nobs}      {\ensuremath{N_{\rm obs}}}
\newcommand{\Nbgd}      {\ensuremath{N_{\rm bgd}}}
\newcommand{\Nul}       {\ensuremath{N_{\rm UL}^{90}}}
\newcommand{\tensev}    {\ensuremath{\times 10^{-7}}}
\def\BRul             {B_{\mathrm{UL}}^{90}}
\newcommand{\dEdM}    {\ensuremath{(\Delta {\rm{M}}, \Delta {\rm E})}}

\graphicspath{.}

\begin{document}

\title{Lepton Flavour Violating $\tau$ decays
\footnote{
\uppercase{T}o appear in the proceedings of 
\uppercase{L}ake \uppercase{L}ouise \uppercase{W}inter \uppercase{I}nstitute 2004 on 
\uppercase{F}undamental \uppercase{I}nteractions (\uppercase{LLWI} 2004), 
\uppercase{L}ake \uppercase{L}ouise, \uppercase{A}lberta, \uppercase{C}anada, 
15-21 \uppercase{F}eb 2004.}}

\author{Swagato Banerjee\footnote{
\uppercase{O}n behalf of the 
{\sl{\uppercase{B}}}{\tiny{\uppercase{A}}}{\sl{\uppercase{B}}}{\tiny{\uppercase{AR}}}
\uppercase{C}ollaboration.}}

\address{University of Victoria\\
PO Box 3055, Victoria, British Columbia, V8W 3P6 Canada.\\
E-mail: swaban@slac.stanford.edu}

\maketitle

\abstracts{
{\bf{Abstract:}}
Searches for lepton flavour violating 
$\tau$ $\ra$ lll and $\tau$ $\ra$ l$\gamma$ decays
at the B-factories are presented. 
Upper limits on the branching ratios are obtained 
$\sim$ $O(10^{-7})$ at 90\% confidence level.}

\section{Introduction}
Lack of any pre-existing experimental evidence 
led the Standard Model (SM) to be built upon the assumption 
that the lepton flavour number is conserved within each family. 
Recent data on neutrino oscillations~[\refcite{neuosc}] have started
to provide hints of lepton flavour ($\lf$) violation in the neutral lepton sector.
Various extensions of the SM to include non-zero neutrino mass
predict manifestations of charged $\lf$ violation 
in terms of branching ratio's of $\tau$ $\ra$ lll (l$\gamma$) decays
at the level of $10^{-14~(40)}$~[\refcite{smlckm}].
However, many beyond the SM processes,
eg.~MSSM + heavy Majorana neutrino with seesaw mechanism~[\refcite{mssm}], 
MSSM with soft SUSY breaking terms~[\refcite{softsusy}],
extra gauge bosons (technicolor)~[\refcite{techni}] 
predict $\lf$ violation at the level of $O$(10$^{-7}$$-$10$^{-10}$)~[\refcite{ma02}].

Present generation $\epem$ B-factories also serve
as $\tau$ factories owing to the large production cross-section of $\tt$ pairs 
at center-of-mass (CM) energy  around the $\Upsilon$(4S) resonance. 
Recent results from the $\tau \ra $ lll search~[\refcite{babar}] in all six possible 
charge conserving decay channels ($\eee$, $\eemw$, $\eemr$, $\emmw$, $\emmr$ and $\mmm$)
with 91.6 $\ifb$ data collected with the $\babar$ experiment is described in detail here.
Results from the $\tau \ra \mu\gamma$ search~[\refcite{belle}] 
with 86.3 $\ifb$ data collected with \textsc{Belle} experiment are also quoted.

\section{$\tau$ $\ra$ lll search}

{\bf{Event Reconstruction:}} 
$\lf$ violating $\tau$ decays are generated according to flat phase space distribution, 
while the other tau of $\tt$ pair produced in $\epem$ annihilation decays according to 
the known branching fractions as modeled by KK2F generator including radiative effects~[\refcite{kkmc}]. 

Events are required to have 4 well reconstructed charged tracks, with each of the 3 tracks 
(from the signal $\tau$ decay) separated by $>$ 90$^{\circ}$ in the CM frame 
from the remaining track (from the other $\tau$ decay).
Pairs of oppositely charged tracks identified as photon conversions in the detector material with
an $\epem$ invariant mass $<$ 30 $\MeV$ are vetoed.
About 50\% of the reconstructed MC signal events pass this 1-3 topology requirement.
\footnote{This estimate includes the 85\% branching fraction for the 1-prong $\tau$ decay.}

Each event is then classified into 6 different decay modes of the signal $\tau$ based upon its charge, 
the ratio of calorimeter energy to track momentum, the ionization loss in the tracking system 
and the shape of the shower in the calorimeter (for electron identification), 
the number of hits in the muon chamber and energy deposits in the calorimeter (for muon identification).
The electron and muon identification have an efficiency per lepton of 91\% and 63\% respectively
for the signal MC.
The mis-identification rate for a hadron selected as an electron or muon in SM 3-prong tau decays 
is 2.2\% or 4.8\% respectively.

{\bf{Background Suppression:}}
The dominant backgrounds for $\eee$, $\emmr$ channels are Bhabhas, for
$\eemr$, $\mmm$ channels are di-muons, and for $\eemw$, $\emmw$ channels are hadronic events ($\qq$)
with the extra tracks coming from un-identified single or multiple conversions.

The QED backgrounds (Bhabha, di-muons) are suppressed by requiring 
the CM momentum of the 1-prong track $<$ 4.8 $\GeV$.
For $\eee$ and $\emmr$ channels, 
the 1-prong track is required not to be an electron, 
while for $\eemr$ and $\mmm$ channels it must not be a muon.
For all these 4 channels, the CM angle $\theta_{13}$ between the 1-prong
momentum and the vector sum of the 3-prong momenta must satisfy $\cos\theta_{13}>-0.9999$, 
and the net transverse momentum of the four tracks $>$ 100 $\MeV$.

To reduce $\qq$ and SM $\tt$ backgrounds, events in the four channels specified above are 
required to have no unassociated calorimeter clusters (photons) in the 3-prong hemisphere 
with energy $>$ 100 $\MeV$ in the laboratory frame, 
while events in all six channels are required to have no track 
in the 3-prong hemisphere that is also consistent with being a kaon.

{\bf{Signal Yield estimation:}}
The distinguishing feature of $\tau \ra$ lll search
is that the signal events cluster around (0,0) 
in (M$_{\mathrm{rec}}$ - m$_{\tau}$, E$^{\star}_{\mathrm{rec}}$ - E$^{\star}_{\mathrm{beam}}$)
$\equiv$ ($\Delta$M, $\Delta$E) plane, where 
M$_{\mathrm{rec}}$ is the reconstructed invariant mass of the three tracks,
m$_{\tau}$ = 1.777 $\GeV$ is the tau mass~[\refcite{bes}],
E$^{\star}_{\mathrm{rec}}$ is the total CM energy of the three tracks
and E$^{\star}_{\mathrm{beam}}$ is the beam CM energy.
Detector effects as well as initial and final state photon radiation 
broaden signal distributions toward lower values of $\Delta$E and $\Delta$M respectively 
as shown in Figure 1.
Signal yield is estimated inside the smaller rectangular boxes shown as inset in the same figure,
chosen so as to give the smallest expected upper limits on the branching fractions in the
background-only hypothesis.

\begin{figure}[b]
 \begin{center}
  \begin{minipage}[l]{2.2in}
   \begin{center}
    \epsfxsize=2.2in\epsfysize=2.6in\epsfbox{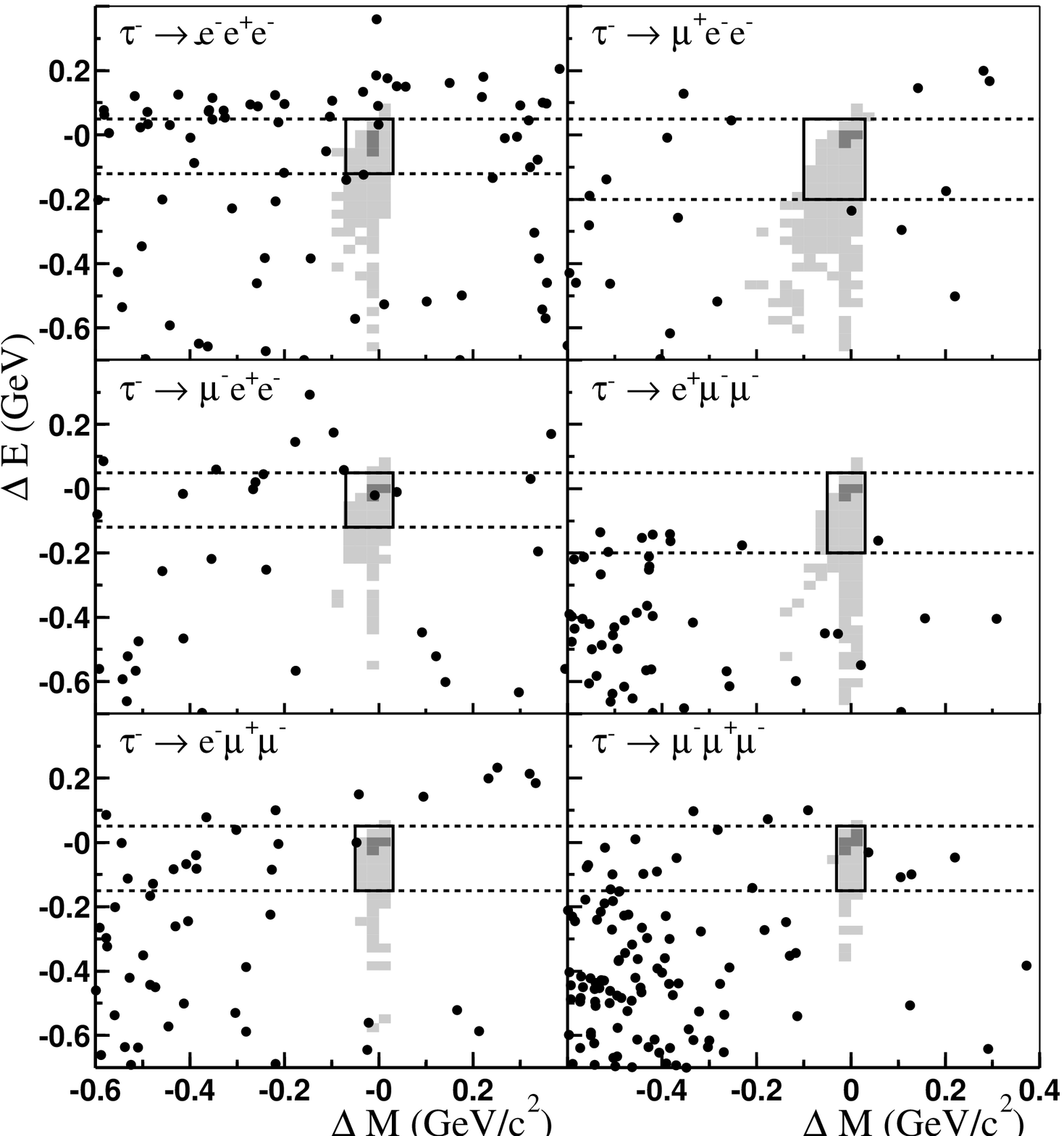}
   \end{center}
\vspace*{-.25cm}
     Figure 1: Observed data as dots and
the boundaries of the signal region for each decay mode.
The dark and light shading indicates contours containing
50\% and 90\% of the selected MC signal events, respectively.
  \end{minipage}
  \begin{minipage}[r]{2.2in}
   \begin{center}
    {\scalebox{.675}{
\begin{tabular}{lccc}
\hline\hline
Decay mode   & \eee         & \eemw        & \eemr \\
\hline\hline
$\varepsilon$ [\%] &$7.3\pm 0.2  $&$11.6\pm 0.4 $&$7.7\pm 0.3$ \\
$\mathrm{Sys}_\varepsilon$ [\%] 
                &        3.3     &   3.7           & 4.1        \\
\hline\hline
$\qq$ bgd.     &$0.67        $&$0.17        $&$0.39$ \\
QED bgd.       &$0.84        $&$0.20        $&$0.23$ \\
$\tt$ bgd.    &$0.00        $&$0.01        $&$0.00$ \\
\hline
\Nbgd           &$1.51\pm 0.11$&$0.37\pm 0.08$&$0.62\pm 0.10$ \\
$\mathrm{Sys}_\mathrm{bgd}$ [\%] 
               &  7.6            &      21        & 16    \\
\hline\hline
\Nobs           &$1           $&$0           $&$1$ \\
\hline
$\BRul$        &$2.0\tensev  $&$1.1\tensev  $&$2.7\tensev$ \\
\hline
\hline
Decay mode      & \emmw        & \emmr        & \mmm \\
\hline\hline
$\varepsilon$ [\%] &$9.8\pm 0.5  $&$6.8\pm 0.4  $&$6.7\pm 0.5$ \\
$\mathrm{Sys}_\varepsilon$ [\%] 
                &   5.0          &       5.3       & 6.8        \\
\hline\hline
$\qq$ bgd.     &$0.20        $&$0.19        $&$0.29$ \\
QED bgd.       &$0.00        $&$0.19        $&$0.01$ \\
$\tt$ bgd.    &$0.01        $&$0.01        $&$0.01$ \\
\hline
\Nbgd           &$0.21\pm 0.07$&$0.39\pm 0.08$&$0.31\pm 0.09$ \\
$\mathrm{Sys}_\mathrm{bgd}$ [\%] 
               &     33         &  21            &   28  \\
\hline\hline
\Nobs           &$0           $&$1           $&$0$ \\
\hline
$\BRul$        &$1.3\tensev  $&$3.3\tensev  $&$1.9\tensev$ \\
\hline
\hline
\end{tabular}
}}
   \end{center}
   Table 1: Efficiency estimates, number of expected background events,
and their relative systematic uncertainties, number of observed events, 
and branching fraction upper limits for each decay mode.
  \end{minipage}
 \end{center}
\end{figure}

The background events have distinctive distributions in the \dEdM\ plane:
$\qq$ events tend to populate the plane uniformly,
while QED backgrounds are restricted to a narrow band at positive values of $\Delta$E 
(owing to the presence of extra charged particles in the event), 
and $\tt$ backgrounds are restricted to negative values of both $\Delta$E and $\Delta$M 
(as the signal topology reconstruction does not account for the missing neutrino(s)).

The expected background rates for each channel are determined 
by fitting a set of probability density functions (PDFs) to the
observed data in the \dEdM\ plane in a grand sideband (GS) region,
defined as the outer boundary of the regions shown in Figure 1 (excluding the signal region).
The analytical forms of the PDFs used are described in detail in Reference~\refcite{babar}.
For both $\qq$ and $\tt$ backgrounds, the shapes of these PDFs are determined
by fits to $\qq$ and $\tt$ background MC samples for each decay mode.
For the QED backgrounds, the shapes of the PDFs are obtained by fitting data control samples 
having a 1-3 topology (without any restriction on $\cos\theta_{13}$), which are enhanced 
in Bhabha or $\mm$ events by requiring that the particle in the 1-prong hemisphere 
is identified as an electron or muon. 

With the shapes of the three background PDFs determined,
an unbinned maximum likelihood fit to the data in the GS region
is performed to obtain the expected number of background events (\Nbgd)
in the signal region, shown along with selection efficiencies ($\varepsilon$) 
and relative systematic uncertainties on $\varepsilon$ 
($\mathrm{Sys}_\varepsilon$) and on \Nbgd ~($\mathrm{Sys}_\mathrm{bgd}$)
in Table 1. The leading contributions to $\mathrm{Sys}_\varepsilon$ come from
statistical precision of data control samples used in particle identification and 
modeling of tracking efficiency, while estimate of $\mathrm{Sys}_\mathrm{bgd}$
is limited by finite data statistics in the GS region used to estimate \Nbgd.

No significant excess is found in any decay mode, and in Table 1 are shown upper limits 
on the branching fractions calculated as $\BRul = \Nul/(2 \varepsilon \mathrm{L} \sigma_{\tau\tau})$, 
where $\Nul$ is the 90\% CL upper limit for the number of signal events when \Nobs\ events are 
observed with \Nbgd\ background events expected for L = 91.6 $\ifb$, $\sigma_{\tau\tau} = 0.89$ nb
\footnote{The luminosity is measured using the observed $\mm$ rate, the cross-sections are estimated
using KK2F~[\refcite{kkmc}] and the uncertainty on the product L $\sigma_{\tau\tau}$ is 2.3\%.}
including all uncertainties~[\refcite{cousins92}]~[\refcite{barlow02}].

\section{Summary}
$\lf$ violating $\tau$ decays have not been observed as yet. 
Upper limits at 90\% confidence level on the branching ratios are obtained:\\
\centerline{$\BRul (\tau \ra \mathrm{lll}) \sim 1-3 \times 10^{-7}$~~[\refcite{babar}],~~
$\BRul (\tau \ra \mathrm{l}\gamma) = 3.1 \times 10^{-7}$~~[\refcite{belle}].}\\
These limits represent an order of magnitude improvement over the previous experimental bounds.

Experimental limits are starting to probe predictions from several theoretical
models with $B$ $\sim$ $O(10^{-7} - 10^{-10})$. 
Improvements over the last few decades on experimental limits on $\BRul$ are shown in Figure 2
along with some theoretical predictions from different LF violating models.

\begin{figure}[!h]
 \begin{center}
  \begin{minipage}[l]{2.2in}
   \begin{center}
    \epsfxsize=2.2in\epsfysize=2.0in\epsfbox{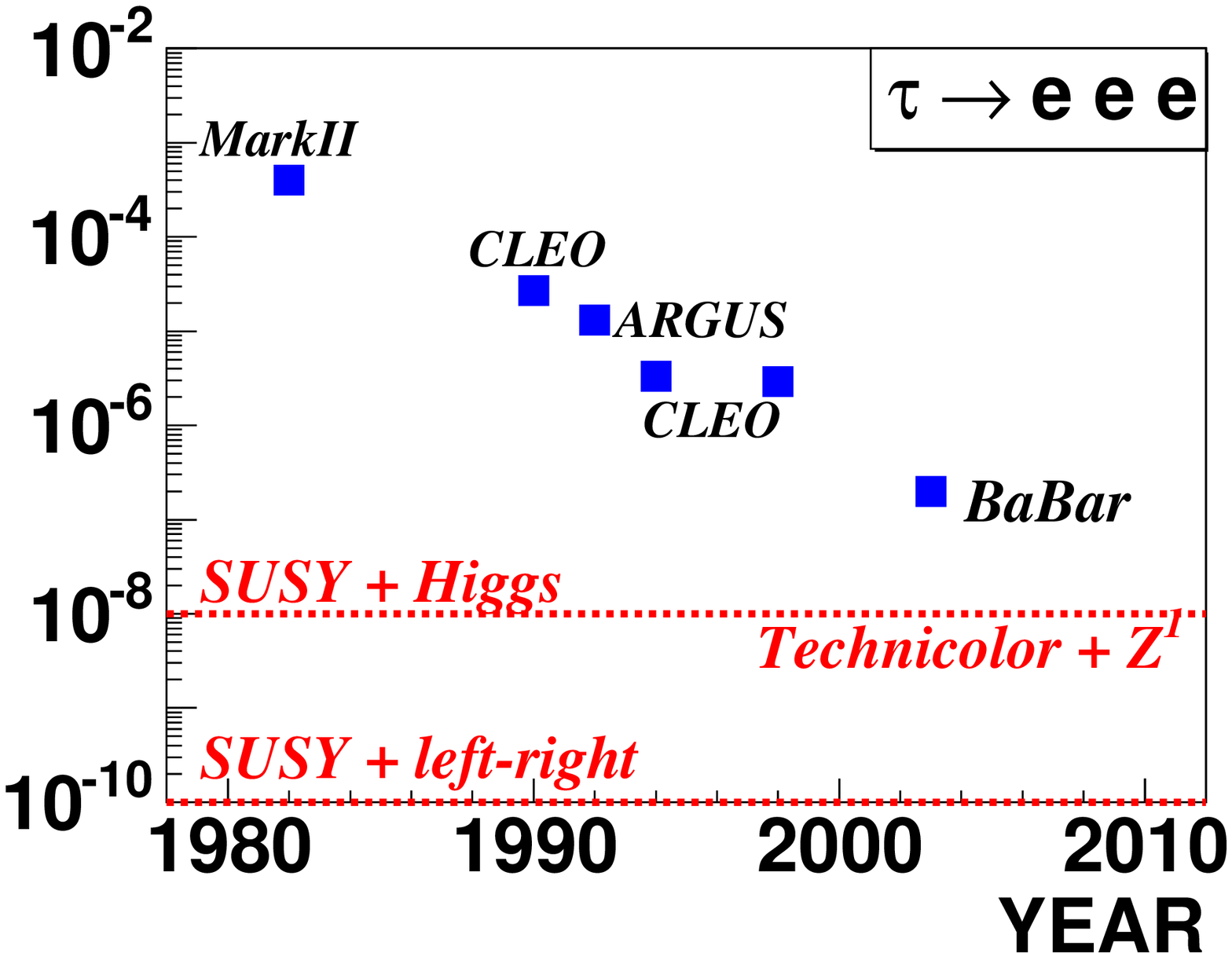}
   \end{center}
  \end{minipage}
  \begin{minipage}[r]{2.2in}
   \begin{center}
    \epsfxsize=2.2in\epsfysize=2.0in\epsfbox{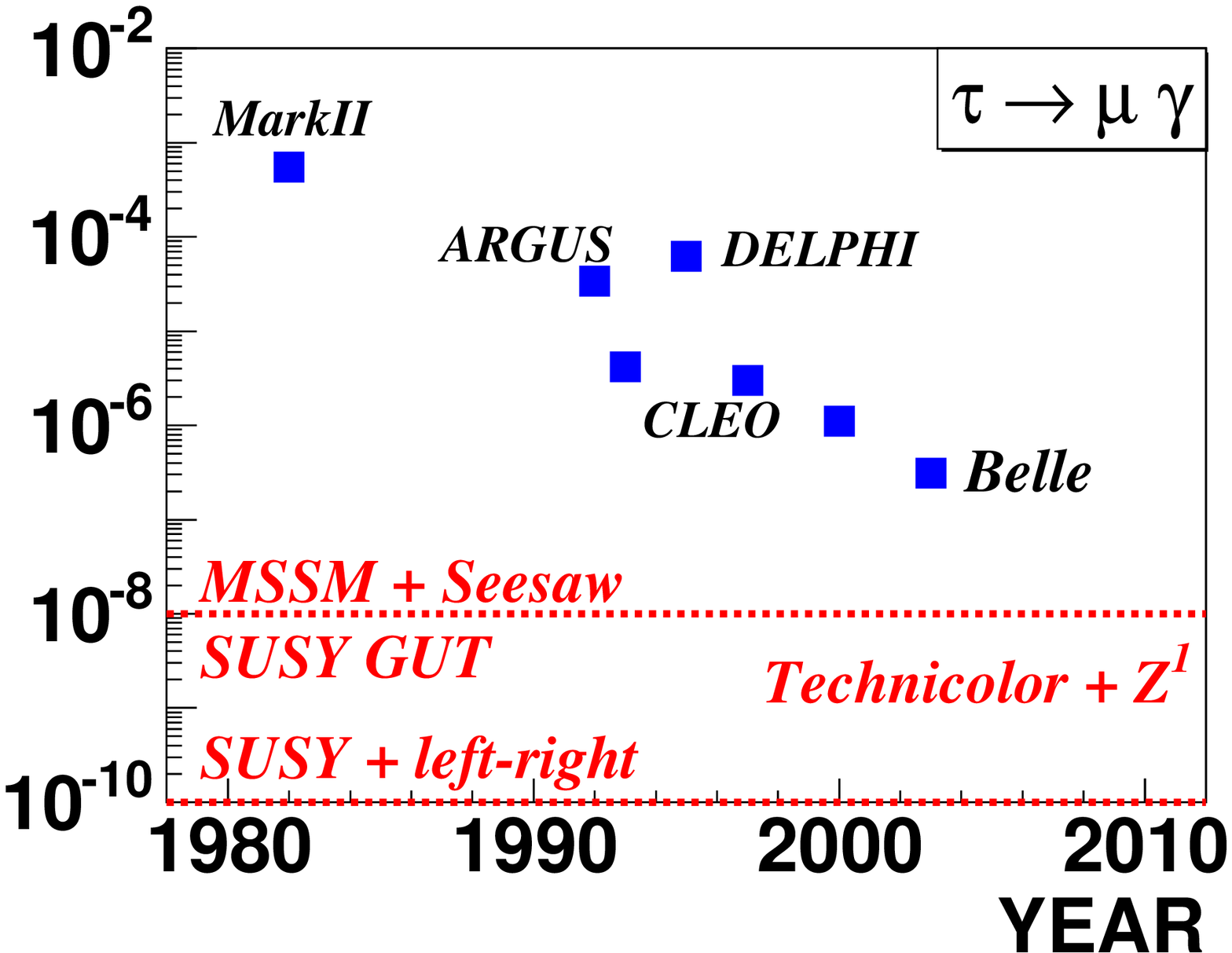}
   \end{center}
  \end{minipage}
\end{center}
\vspace*{-.25cm}
Figure 2: Evolution of experimental bounds ($\BRul$) and some predictions.
\end{figure}


\vspace*{-.75cm}

\begin{thebibliography}{0}

\bibitem{neuosc}
K2K Collaboration, M.~H.~Ahn {\it et al.}, Phys.\ Rev.\ Lett. {\bf 90}, 041801 (2003);
KamLAND Collaboration, K.~Eguchi {\it et al.}, Phys.\ Rev.\ Lett. {\bf 90}, 021802 (2003);
SNO Collaboration, Q.~R.~Ahmad {\it et al.}, Phys.\ Rev.\ Lett. {\bf 89}, 011301 (2002);
Super-Kamiokande Collaboration, Y.~Fukuda {\it et al.}, Phys.\ Rev.\ Lett. {\bf 81}, 1562 (1998).

\bibitem{smlckm} 
 Xuan-Yem Pham,  Eur.\ Phys.\ J.\ {\bf C8}, 513-516 (1999).

\bibitem{mssm} 
S.~T.~Petcov, S.~Profumo, Y.~Takanishi, C.~E.~Yaguna,
Nucl.\ Phys.\ {\bf B676}, 453 (2004);
J.~R.~Ellis, J.~Hisano, M.~Raidal, Y.~Shimizu,
Phys.\ Rev.\ {\bf D66}, 115013 (2002).

\bibitem{softsusy} 
J.~R.~Ellis, M.~E.~Gomez, G.~K.~Leontaris, S.~Lola, D.~V.~Nanopoulos,
Eur.\ Phys.\ J.\ {\bf C14}, 319 (2000);
T.~Fu.~Feng, T.~Huang, X.~Q.~Li, X.~M.~Zhang, S.~M.~Zhao,
Phys.\ Rev.\ {D68}, 016004 (2003).

\bibitem{techni} 
C.~X.~Yue, Y.~M.~Zhang, L.~J.~Liu,
Phys.\ Lett.\ {\bf B547}, 252 (2002).

\bibitem{ma02}
E.~Ma,
Nucl.\ Phys.\ B~Proc.\ Suppl.\ {\bf 123}, 125 (2003).

\bibitem{babar} 
BaBar Collaboration, B.~Aubert {\it et at.},
Phys.\ Rev.\ Lett. {\bf 92}, 121801 (2004).

\bibitem{belle}
Belle Collaboration, K.~Abe {\it et al.},
hep-ex/0310029.

\bibitem{kkmc}B.~F.~Ward, S.~Jadach, and Z.~Was,
Nucl.\ Phys.\ Proc.\ Suppl.\  {\bf 116}, 73 (2003);
S.~Jadach, Z.~Was, R.~Decker, and J.~H.~Kuhn,
Comput.\ Phys.\ Commun.\  {\bf 76}, 361 (1993);
E.~Barberio and Z.~Was,
Comput.\ Phys.\ Commun.\  {\bf 79}, 291 (1994).


\bibitem{bes}
BES Collaboration, J.~Z.~Bai {\it et al.}, Phys.\ Rev. {\bf D53}, 20 (1996).

%
%
\bibitem{cousins92}
R.~D.~Cousins and V.~L.~Highland,
Nucl.\ Instrum.\ Meth.\ A {\bf 320}, 331 (1992).

\bibitem{barlow02}
R.~Barlow,
Comput.\ Phys.\ Commun.\  {\bf 149}, 97 (2002).


\end{thebibliography}
\end{document}